\documentclass[a4paper,11pt]{article}
\usepackage{jcappub}
\usepackage{amsmath}
\usepackage{graphicx}
\usepackage{xcolor}
\usepackage{placeins}

\newcommand{\rmd}{{\rm d}}
\newcommand{\ukarcmin}{\ensuremath{\mu {\rm K{\text -}arcmin}}}

\begin{document}
\title{Robust Measurements of the Cosmic Microwave Background Lensing Power Spectrum in the Presence of Anisotropic Cosmic Birefringence}

\author[a]{Rui Zhou,}
\author[a]{Michael Doohan,}
\author[a]{and Christian L. Reichardt}

\affiliation[a]{School of Physics, University of Melbourne, Parkville, VIC 3010, Australia}

\emailAdd{rui.zhou@student.unimelb.edu.au}

\abstract{
Achieving the scientific potential of upcoming high-sensitivity cosmic microwave background (CMB) polarization surveys requires lensing analyses that are robust to unmodeled distortions of the polarization field.
One such possible distortion is anisotropic cosmic birefringence, a direction-dependent rotation of the linear polarization plane that may arise from parity-violating physics such as axion-like fields.
We quantify the impact of both modeled and unmodeled anisotropic birefringence on the CMB lensing measurement in the Marginal Unbiased Score Expansion (MUSE) framework.
Considering noise levels expected for current and future CMB surveys, we find that neglecting birefringence produces a positive bias in the lensing bandpowers that increases towards small angular scales.
For a scale-invariant birefringence power spectrum $L(L+1)C_L^{\alpha\alpha}/2\pi=10^{-5}\,{\rm rad}^2$, the positive fractional bias can reach $(10.1\pm1.1)\%$ of the lensing power at $1375<L<2000$.
Jointly estimating lensing and birefringence in the MUSE framework recovers the input lensing bandpowers and birefringence amplitude without detectable bias across all tested configurations.
The joint MUSE estimator also reduces the scatter of the recovered birefringence amplitude by factors of $1.6\pm0.2$ to $2.4\pm0.3$ relative to a joint quadratic estimator, with the largest gains in the highest signal-to-noise configurations.
These results indicate that a joint-field reconstruction with MUSE can recover unbiased estimates of the lensing and birefringence power spectra while still improving precision relative to quadratic estimators.}

\maketitle

\section{Introduction}
\label{sec:introduction}

Gravitational lensing by intervening large-scale structure remaps the temperature and polarization anisotropies of the cosmic microwave background (CMB), generating lensing-induced $B$-mode polarization and characteristic non-Gaussian correlations between CMB modes~\citep{Seljak1996Lensing,Zaldarriaga1998LensingPolarization,Lewis2006Weak}.
These correlations allow the lensing potential to be reconstructed, turning CMB polarization into a probe of the projected matter distribution and the growth of cosmic structure~\citep{Seljak1999DarkMatter,Hu2001Mapping,Hu2002Mass,Okamoto2003Full}.
Current and forthcoming high-sensitivity experiments, including SPT-3G, BICEP3/Keck Array, and the Simons Observatory, are measuring or will measure polarization with substantially lower noise than existing experiments~\citep{Benson2014SPT3G,Grayson2016BICEP3,SimonsObservatory2019}.
The resulting increase in lensing sensitivity makes it important to control distortions not included in the reconstruction model.

One such distortion could be the hypothesized cosmic birefringence, which would rotate the plane of linear polarization as CMB photons propagate to us from the last-scattering surface.
This rotation would arise from parity-violating physics, such as an axion-like field coupled to electromagnetism through a Chern--Simons interaction~\citep{Carroll1990Limits,Carroll1991Einstein}, or from Faraday rotation by primordial magnetic fields~\citep{Kosowsky1996Faraday,Scoccola2004Faraday}.
Spatial variations in the axion-like field would produce spatial variations in the rotation angle, known as anisotropic birefringence~\citep{Yadav2009Constraining,Gluscevic2009Derotation}.
Similar to lensing, such a rotation would induce statistical anisotropy and mode coupling in the observed CMB maps, although its response functions and parity properties differ~\citep{Kamionkowski2009Derotate,Cai2023Impact}.
Reconstructions of the rotation field have so far yielded only upper limits on the amplitude of anisotropic birefringence~\citep{Gluscevic2012First,Namikawa2020ACT,Bianchini2020Searching,Kramer2026ACT}.
Nevertheless, a rotation signal consistent with these limits can bias the reconstructed lensing power spectrum if neglected in the analysis~\citep{Cai2023Impact}.

The standard way to reconstruct lensing and other distortion fields is through quadratic estimators (QEs), which are computationally efficient and exploit the mode coupling that lensing induces between different CMB modes~\citep{Hu2002Mass,Okamoto2003Full,Kamionkowski2009Derotate}.
Within this framework, Cai et al.~\citep{Cai2023Impact} showed that the lensing and birefringence responses are orthogonal at leading order, so that birefringence does not bias the reconstructed lensing field.
The lensing power spectrum, however, is still biased through connected four-point terms.
Using the $EB$ estimator, Cai et al.~\citep{Cai2023Impact} found that this bias grows with $L$ and scales linearly with the birefringence amplitude.
A follow-up identified this bias as an $N^{(1)}$-like term and provided estimators for it~\citep{Cai2024Efficient}, but left a detailed analysis of mitigation to future work.

This work presents such a mitigation strategy, jointly modeling the lensing and birefringence fields in a common forward model.
Likelihood-based reconstruction offers a route beyond the quadratic approximation by maximizing the posterior of the distortion fields~\citep{Hirata2003Reconstruction,Carron2017Maximum}.
However, this framework targets the fields themselves rather than the lensing and birefringence power spectra.
Inferring these spectra instead means integrating the fields out of the likelihood.
Because the latent fields are high-dimensional and their joint posterior is non-Gaussian, this integral is computationally expensive for large datasets~\citep{Anderes2015Bayesian,Millea2019Bayesian,Millea2020Sampling}.
To overcome this, Marginal Unbiased Score Expansion (MUSE) approximates the marginal score by evaluating the parameter score at a maximum-a-posteriori (MAP) latent-field solution and calibrating it with simulations drawn from the model, rather than carrying out the full marginalization~\citep{Millea2022MUSE}.
Since its introduction, MUSE has been applied to joint CMB and lensing-spectrum inference, patchy-reionization and lensing inference, and tests of non-Gaussian extragalactic-foreground contamination~\citep{Millea2022MUSE,Bianchini2023Patchy,Doohan2025Quantifying}.
It has also been applied to SPT-3G polarization data to measure the lensing and delensed $E$-mode power spectra~\citep{Ge2025Cosmology}.
Since the hierarchical formulation of MUSE can accommodate additional latent distortion fields, the method can be extended to anisotropic cosmic birefringence.
This is the first time MUSE has been extended to jointly infer the amplitude of a scale-invariant anisotropic birefringence spectrum and the CMB lensing bandpowers.

The closest precedent that includes birefringence is the joint MAP framework of Darwish~\citep{Darwish2025Optimal}, which reconstructs the lensing, birefringence, and patchy-screening fields together.
More recently, Belkner and Carron~\citep{Belkner2026Delensing} jointly reconstructed the lensing and birefringence fields at the MAP point to constrain the birefringence amplitude.
Neither work, however, examines the magnitude of the birefringence-induced bias on the recovered lensing power spectrum.

Using flat-sky simulations, we quantify the bias in the MUSE lensing power spectrum when birefringence is present in the maps but omitted from the analysis.
We show that the birefringence-induced bias in the lensing power spectrum is effectively mitigated by the joint-field MUSE framework.
We show that the joint MUSE framework returns more precise measurements of the lensing and birefringence power spectra than the joint QE.

The remainder of this paper is organized as follows.
Section~\ref{lensing_birefringence_and_bandpowers} reviews CMB lensing and anisotropic cosmic birefringence before introducing the joint forward model.
Section~\ref{pipeline} describes the simulations on which we compare three estimators: joint MUSE, lensing-only MUSE, and joint QE.
Section~\ref{quadratic_estimators} describes the baseline joint $EB$ QE, including its bias subtraction, while Section~\ref{muse_approximation} presents the joint MUSE estimator.
Section~\ref{results} presents our results, and Section~\ref{conclusions} summarizes our findings and outlines future work.

\section{CMB Polarization Distortions}
\label{lensing_birefringence_and_bandpowers}

Both lensing and birefringence distort the polarization pattern from last scattering, with lensing locally translating the pattern and birefringence locally rotating the polarization direction.
We review the two effects in turn, before presenting the joint forward model used throughout this work.

\subsection{CMB gravitational lensing}

The deflection due to gravitational lensing can be expressed as the gradient of the lensing potential $\phi$, a line-of-sight projection of the gravitational potential between the observer and last scattering.
The lensing potential traces the integrated matter distribution over a broad range of redshifts, and the lensing potential power spectrum probes cosmological geometry and the growth of large-scale structure~\citep{Seljak1999DarkMatter,Hu2001Mapping,Das2011Detection,Planck2018Lensing,Madhavacheril2024ACTLensing}.
In the Born approximation, the lensed temperature and polarization fields are related to their unlensed counterparts by~\citep{Seljak1996Lensing,Hu2000Harmonic,Lewis2006Weak,Namikawa2014Cosmology}
\begin{equation}
    \widetilde{X}(\hat n)=X\big(\hat n+\nabla\phi(\hat n)\big),\qquad X\in\{T,Q,U\},
    \label{eq:lensing_remapping}
\end{equation}
where $\nabla\phi$ is the deflection field, $T$ is the temperature, and $Q$ and $U$ are the Stokes parameters of linear polarization.
The polarization field can be decomposed into gradient-like $E$ modes, which are produced by scalar perturbations, and curl-like $B$ modes, which are not produced by scalar perturbations at last scattering~\citep{Kamionkowski1997Statistics,Zaldarriaga1997AllSky}.
The remapping due to lensing converts a fraction of the $E$ modes into $B$ modes and generates non-Gaussian correlations between CMB modes~\citep{Zaldarriaga1998LensingPolarization,Hu2000Harmonic,Lewis2006Weak,Hanson2013BMode,Polarbear2014BMode}.
We model $\phi$ as a statistically isotropic Gaussian field, so that its power spectrum $C_L^{\phi\phi}$ fully specifies it.
We therefore infer $C_L^{\phi\phi}$, in bandpowers, as the primary lensing quantity of this work.

\subsection{Anisotropic cosmic birefringence}
\label{sec:cosmic_birefringence}

In the simplest models, cosmic birefringence arises from a pseudoscalar field $a$ coupled to electromagnetism through a Chern--Simons interaction, $\mathcal{L}_{\mathrm{CS}}=(g_{a\gamma}/4)\,a\,F_{\mu\nu}\widetilde{F}^{\mu\nu}$, where $g_{a\gamma}$ is the axion--photon coupling, $F_{\mu\nu}$ is the electromagnetic field-strength tensor, and $\widetilde{F}^{\mu\nu}$ is its dual.
A photon's linear polarization is then rotated by an angle $\alpha(\hat n)=g_{a\gamma}\Delta a(\hat n)/2$~\citep{Carroll1990Limits,Carroll1991Einstein,Diego1992Effects,Komatsu2022New}, where $\Delta a(\hat n)$ denotes the change in the pseudoscalar field along the line of sight.
Mathematically, this rotation acts on the spin-2 polarization combinations $P_\pm\equiv Q\pm iU$ as
\begin{equation}
    P'_\pm(\hat n)=e^{\pm2i\alpha(\hat n)}P_\pm(\hat n).
    \label{eq:rotation}
\end{equation}

The rotation can be decomposed into isotropic and anisotropic components.
Homogeneous evolution of $a$ would give an isotropic rotation, which we set to zero throughout this work.
Spatial fluctuations of $a$ would yield an anisotropic rotation, inducing off-diagonal correlations between CMB polarization modes~\citep{Kamionkowski2009Derotate,Yadav2009Constraining,Gluscevic2009Derotation}.
If $a$ is light during inflation, its fluctuations are nearly Gaussian and scale invariant on large scales ($L\lesssim100$), and so is the resulting rotation field~\citep{Yadav2009Constraining,Caldwell2011Cross,Cai2023Impact}.
Motivated by this, we model $\alpha$ as a Gaussian field with an exactly scale-invariant spectrum,
\begin{equation}
    \frac{L(L+1)}{2\pi}C_L^{\alpha\alpha}=A_{\rm CB} ~[{\rm rad}^2],
\end{equation}
where $A_{\rm CB}$, in units of rad$^2$, denotes the amplitude of the birefringence power spectrum.
Currently, the tightest limits come from SPT-3G with $A_{\rm CB}<2.7\times10^{-6}$~\citep{coerver26} and BICEP/Keck, with $A_{\rm CB}<4.4\times10^{-6}$ (95\% C.L.)~\citep{BICEPKeck2023XVII}.

\subsection{Joint forward model}
\label{sec:forward_model}

The relative ordering of lensing and anisotropic birefringence depends on the physical origin and time evolution of the rotation field, with both orderings adopted in previous studies~\citep{Cai2023Impact,Darwish2025Optimal}.
For simplicity, we consider an early-time rotation field, with lensing acting on the already-rotated polarization field~\citep{Darwish2025Optimal}.
Applying Eq.~\eqref{eq:rotation} and then Eq.~\eqref{eq:lensing_remapping} gives
\begin{equation}
    \label{model}
    \widetilde{P}_\pm(\hat n)=e^{\pm 2i\alpha(\hat n+\nabla\phi(\hat n))}\,P_\pm\big(\hat n+\nabla\phi(\hat n)\big),
\end{equation}
in which $\alpha$ is evaluated at the deflected position $\hat n+\nabla\phi(\hat n)$.
Equation~\eqref{model} defines the forward model used consistently in our simulations and likelihood evaluation.

\section{Simulations}
\label{pipeline}

To quantify the lensing bias from leaving birefringence out of the model, and to compare the precision of the estimators, we test them on flat-sky simulations with known input lensing and birefringence fields.
We apply all three estimators described in Sections~\ref{quadratic_estimators} and~\ref{muse_approximation} (joint MUSE, lensing-only MUSE, and joint QE) to the same simulated skies, so that differences between them are not driven by sample variance.
All three use only the polarization maps, as in the MUSE analysis of SPT-3G data~\citep{Ge2025Cosmology}.

The fiducial unlensed CMB and lensing-potential power spectra are computed with \textsc{camb}~\citep{Lewis2000CAMB} for a tensor-to-scalar ratio $r=10^{-3}$.
We draw Gaussian realizations of the unlensed CMB sky, the lensing potential $\phi$, and the rotation angle $\alpha$ on $256\times256$ flat-sky maps with $3'$ pixels ($\simeq164\,\mathrm{deg}^2$) and no mask.
For each realization, we rotate the polarization maps by the birefringence field and then lens the rotated maps following Eq.~\eqref{model}~\citep{Darwish2025Optimal}, as in the inference model.
We then smooth the maps with a $1'$ beam and bandpass-filter them to $20<\ell<5000$.
Finally, we add Gaussian noise to the polarization maps with power spectrum
\begin{equation}
    N_\ell = \Delta_P^2 \left[ 1 + \left( \frac{\ell_{\rm knee}}{\ell} \right)^{\gamma_{\rm knee}} \right],
\end{equation}
consisting of white noise at level $\Delta_P$ and a $1/f$ component that rises at low $\ell$, with $\ell_{\rm knee}=100$ and $\gamma_{\rm knee}=3$.
We consider two noise levels, $\Delta_P=1.4$ and $2.3\,\mu$K-arcmin, as in Ref.~\citep{Darwish2025Optimal}.
The noise levels represent CMB-S4-like~\citep{CMBS4ScienceBook2016} and SPT-3G-like surveys, with the latter matching the projected depth of the $1500\,\mathrm{deg}^2$ SPT-3G Main field~\citep{Prabhu2024Testing}.
For each noise level, we simulate three input birefringence amplitudes, $A_{\rm CB}^{\rm input}= 4.4\times10^{-6}$ (near current limits), $10^{-5}$ (slightly above current limits), and $10^{-4}$ (larger than current limits).
The two larger amplitudes allow us to characterize the lensing bias and test the joint analysis in a strong-signal regime, while the smallest\footnote{Our simulations are at the BICEP/Keck limit instead of the $2.7\times 10^{-6}$ limit from SPT-3G because we ran the simulations before the SPT-3G measurement was released.} matches the BICEP/Keck upper limit~\citep{BICEPKeck2023XVII}.
This gives six configurations, each with 50 independent realizations.
We also simulate a no-birefringence case on the same realizations, with the input rotation field set to zero.

\section{Quadratic-estimator reconstruction}
\label{quadratic_estimators}

As a baseline for MUSE, we reconstruct $\phi$ and $\alpha$ with $EB$ quadratic estimators (QEs)~\citep{Hu2002Mass,Okamoto2003Full}.
Both distortions induce correlations between $E$ and $B$ modes, which, to linear order in $\phi$ and $\alpha$, can be written as
\begin{equation}
    \langle E(\boldsymbol{\ell}_1)B(\boldsymbol{\ell}_2)\rangle
    =f^\phi(\boldsymbol{\ell}_1,\boldsymbol{\ell}_2)\,\phi(\boldsymbol{L})
    +f^\alpha(\boldsymbol{\ell}_1,\boldsymbol{\ell}_2)\,\alpha(\boldsymbol{L}),
    \label{eq:qe_linear_response}
\end{equation}
with $\boldsymbol{L}=\boldsymbol{\ell}_1+\boldsymbol{\ell}_2$, $f^\phi=(C^{EE}_{\ell_1}\boldsymbol{L}\cdot\boldsymbol{\ell}_1-C^{BB}_{\ell_2}\boldsymbol{L}\cdot\boldsymbol{\ell}_2)\sin2\varphi_{12}$, and $f^\alpha=2(C^{EE}_{\ell_1}-C^{BB}_{\ell_2})\cos2\varphi_{12}$, where $\varphi_{12}$ is the angle between $\boldsymbol{\ell}_1$ and $\boldsymbol{\ell}_2$ and the spectra are lensed~\citep{Hu2002Mass,Okamoto2003Full,Kamionkowski2009Derotate,Yadav2009Constraining,Gluscevic2009Derotation,Winston2022Probing}.

Each QE is a weighted product of the observed modes,
\begin{equation}
    \hat x(\boldsymbol{L})
    =A^x_L\int\frac{d^2\boldsymbol{\ell}}{(2\pi)^2}\,
    \frac{f^x(\boldsymbol{\ell},\boldsymbol{L}-\boldsymbol{\ell})}{C^{EE}_{{\rm tot},\ell}\,C^{BB}_{{\rm tot},|\boldsymbol{L}-\boldsymbol{\ell}|}}\,
    E(\boldsymbol{\ell})\,B(\boldsymbol{L}-\boldsymbol{\ell}),
    \qquad x\in\{\phi,\alpha\},
    \label{eq:qe}
\end{equation}
where $C_{\rm tot}$ is the lensed spectrum plus noise and $A^x_L$ normalizes the estimator to unit response~\citep{Hu2002Mass,Yadav2009Constraining}.
Because $f^\phi\propto\sin2\varphi_{12}$ and $f^\alpha\propto\cos2\varphi_{12}$, the two estimators do not respond to each other's field at leading order~\citep{Kamionkowski2009Derotate,Cai2023Impact}, allowing each to be normalized independently.
Before computing the power spectra, we divide each reconstructed field by its residual response, the ratio of its cross-spectrum with the input field to the input power.
We estimate this response from 100 simulations.

Several biases must be subtracted from the raw QE power spectra to obtain unbiased estimates.
The resulting debiased spectrum is given by~\citep{Cai2023Impact}
\begin{equation}
    \hat C^{xx}_L=C^{\hat x\hat x}_L-N^{(0),x}_L-N^{(1),x}_L-N^{{\rm MC},x}_L,\qquad x\in\{\phi,\alpha\}.
\end{equation}
The Gaussian bias $N^{(0),x}_L$ arises from the disconnected part of the CMB four-point function and is estimated with the realization-dependent method~\citep{Hanson2011LensingPower,Namikawa2013BiasHardened}, which combines the observed maps with 100 pairs of Gaussian simulations of the lensed CMB without $\phi$ or $\alpha$.
Sourced by both lensing and rotation, the connected bias $N^{(1),x}_L$ is the leading connected contribution beyond the signal.
As in Ref.~\citep{Cai2024Efficient}, we estimate it from 100 sets of simulation pairs with different primary CMB and noise realizations: pairs with the same lensing field but different rotation fields and pairs with the same rotation field but different lensing fields, each minus pairs with different realizations of both fields.
The Monte Carlo correction $N^{{\rm MC},x}_L$ represents the remaining biases not accounted for by $N^{(0),x}_L$ and $N^{(1),x}_L$, such as higher-order reconstruction noise~\citep{Cai2023Impact}.
We estimate it from 100 simulations as the mean difference between the spectrum after $N^{(0),x}_L$ and $N^{(1),x}_L$ subtraction and the input power.
All simulations used for the response and bias terms are independent of the 50 test realizations.
Apart from the Gaussian simulations used for $N^{(0),x}_L$, they contain both $\phi$ and $\alpha$, with $\alpha$ drawn at the input $A_{\rm CB}$ of each configuration, and we therefore refer to this estimator as the joint QE.

Finally, to obtain a single amplitude $A_{\rm CB}$ for comparison with the MUSE estimate presented below, we fit the scale-invariant spectrum to the eight rotation bandpowers over $30\leq L\leq1000$.
This fit is performed by minimizing $\chi^2$, using a bandpower covariance matrix estimated from 100 independent simulations.

\section{MUSE inference}
\label{muse_approximation}

Exact marginal maximum-likelihood inference requires marginalizing the likelihood over high-dimensional latent fields, which is computationally expensive for large datasets.
With the marginal score defined as $s_i^{\rm marginal}(\theta,d)\equiv \frac{\rmd}{\rmd\theta_i}\log\mathcal{P}(d\,|\,\theta)$, the marginal maximum-likelihood estimate satisfies $s_i^{\rm marginal}(\hat\theta^{\rm MMLE},d)=0$.
MUSE approximates this score at the MAP solution for the latent fields and calibrates the approximation with simulations drawn from the model~\citep{Millea2022MUSE}.
This approximation leads to a much cheaper approach than exact marginalization, while the resulting estimate remains asymptotically unbiased and near-optimal.

Specifically, for fixed $\theta$, let $z_J(\theta,d)\equiv\operatorname*{arg\,max}_{z}\log\mathcal{P}(d,z\,|\,\theta)$ denote the MAP solution for the latent fields.
The score of the joint probability evaluated at this solution is $s_i^{\rm MAP}(\theta,d)\equiv \frac{\rmd}{\rmd\theta_i}\log\mathcal{P}\!\left(d,z_J(\theta,d)\,|\,\theta\right)$.
MUSE removes the bias introduced by replacing marginalization with MAP evaluation by subtracting the corresponding mean score over simulations drawn from the model at the same $\theta$,
\begin{equation}\label{eq:muse_score}
    s_i^{\rm MUSE}(\theta,d)
    \equiv
    s_i^{\rm MAP}(\theta,d)
    -
    \left\langle
    s_i^{\rm MAP}(\theta,d')
    \right\rangle_{d' \sim \mathcal{P}(d'\,|\,\theta)}.
\end{equation}
The MUSE estimate $\hat\theta$ is the root of $s^{\rm MUSE}(\theta,d)=0$, which we solve with the pseudo-Newton--Raphson method,
\begin{equation}\label{eq:muse_update}
    \hat{\theta}_{n+1}
    =
    \hat{\theta}_n
    -
    \beta K^{-1}
    s^{\rm MUSE}(\hat{\theta}_n,d),
\end{equation}
where $\beta$ is a step size and $K$ is a preconditioning matrix.
We estimate the uncertainties on $\hat\theta$ reported for the ensemble comparisons directly from the empirical scatter across the 50 simulated realizations of each configuration, and have verified that the observed scatter agrees well with that obtained from the formal MUSE covariance~\citep{Millea2022MUSE}.

For joint lensing--birefringence inference, the latent state, parameter vector, and lensing bandpowers become
\begin{equation}
    z=(f,\phi,\alpha),
    \qquad
    \theta=\bigl(A_{\phi}^{1},\ldots,A_{\phi}^{N_{\phi}},A_{\rm CB}\bigr)^{T},
    \qquad
    C_L^{\phi\phi}(\theta)=A_\phi^b C_L^{\phi\phi,\mathrm{fid}}\ \ (L\in b),
    \label{eq:lensing_bandpower_definition}
\end{equation}
where $f$ is the unlensed CMB polarization field, $A_\phi^b=1$ corresponds to the fiducial lensing power in bin $b$, and the $N_\phi=8$ bins have equal width in $\ln L$ over $100\leq L\leq2000$.
The birefringence amplitude $A_{\rm CB}$ is inferred from rotation modes with $30\leq L\leq1000$.

We use a modified version of \textsc{CMBLensing.jl}\footnote{\url{https://github.com/rzhou-physics/CMBLensingBirefringence.jl}, a fork of \url{https://github.com/marius311/CMBLensing.jl}.} to perform the map-level operations and solve the MAP problems.
This differs from the lensing-only MUSE branch~\citep{Millea2022MUSE} only in adding $\alpha$ as a latent field to the model.
At each MUSE iteration, we generate 64 simulations at the current parameter values and solve the MAP problem for both the target map and every simulated map.
The difference of their MAP scores gives the debiased score of Eq.~\eqref{eq:muse_score}, which we use to update the lensing bandpowers and $A_{\rm CB}$ simultaneously via Eq.~\eqref{eq:muse_update}.
We take $K=-\mathrm{diag}(J)$, with $J$ the covariance of the MUSE score over the simulations, and a step size $\beta=1/[1+0.03(n-1)]$ that decreases with iteration $n$ as the solution is approached.
We iterate until convergence, for at most 50 iterations.
For computational efficiency, we warm-start each latent-field optimization from the solution obtained at the previous MUSE iteration, separately for the observed data and each simulated dataset, and reuse the same random seeds across iterations, so that the simulation noise in the MUSE score does not change from one iteration to the next.
Since $\theta$ changes little between iterations, the warm start substantially reduces the number of optimization steps required.

\section{Results}
\label{results}

We now look at the bias in and precision of estimates of the lensing and rotation power spectra.
We show that for simulated maps without birefringence, jointly modeling the rotation field has minimal effect on the lensing measurement.
We then quantify the bias incurred from neglecting birefringence when measuring CMB lensing.
We show that the joint MUSE approach effectively mitigates this bias, with a $\lesssim2\%$ increase in the lensing bandpower uncertainties.
Finally, we check the quality of the birefringence estimation.

\subsection{No-birefringence case}
\label{no_birefringence}

We first test the case with no birefringence, which serves as a baseline for the cases with birefringence.
In this case, joint MUSE recovers the lensing bandpowers without detectable bias, as does lensing-only MUSE, and modeling the rotation field changes the single-realization scatter of the lensing bandpowers by less than $1\%$ in every bin.
When no birefringence is present, marginalizing over birefringence in the MUSE model has no impact on the CMB lensing measurement (outside of a small computational cost).

At polarization noise levels of $\Delta_P=1.4$ and $2.3\,\mu$K-arcmin, the joint QE gives $\hat A_{\rm CB}=(0.06\pm0.19)\times10^{-6}$ and $(0.20\pm0.24)\times10^{-6}$, respectively, both consistent with zero.
Because the joint MUSE update imposes a positive condition on $A_{\rm CB}$ (specifically that it is at least $10^{-4}$ times the weakest nonzero input amplitude), the MUSE estimates do not scatter around zero.
The posterior peaks at the lower edge of the $A_{\rm CB}$ prior for both noise levels.
The 95\% upper limits are $A_{\rm CB}<5.6\times10^{-7}$ and $1.1\times10^{-6}$, respectively, for the two noise levels, consistent with the expected $1.6\,\sigma$ uncertainties on $A_{\rm CB}$.
There is no evidence for a spurious detection of birefringence power.

As expected at these noise levels, the MUSE estimator improves upon the precision of the QE.
The uncertainty on the joint QE lensing bandpowers is $1.51\pm0.12$ to $4.91\pm0.65$ times that of joint MUSE bandpowers, with the improvement increasing towards high $L$.

\subsection{Birefringence-induced lensing bias}
\label{lensing_bias}

\begin{figure}[!htbp]
    \centering
    \includegraphics[width=0.90\linewidth]{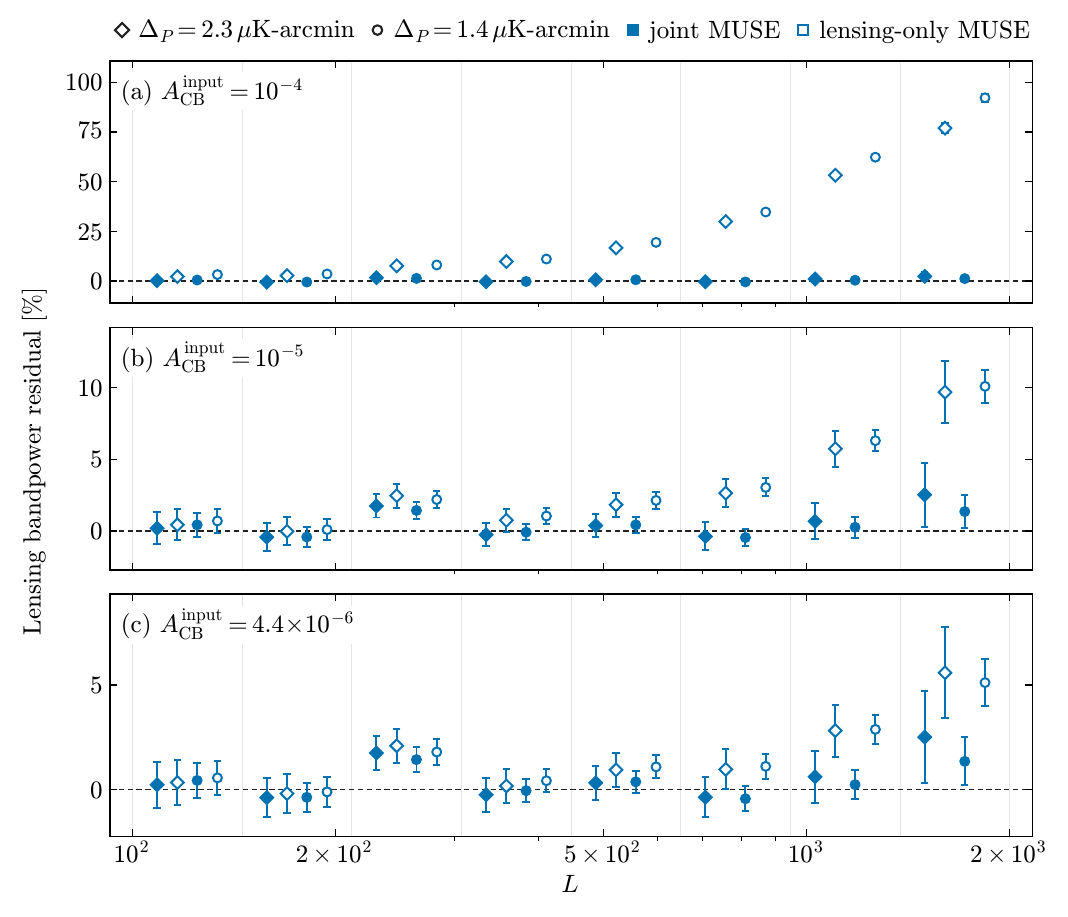}
    \caption{When estimating the lensing bandpowers, lensing-only MUSE (open) is biased high by an amount that grows with $A_{\rm CB}^{\rm input}$ and with $L$, while joint MUSE (filled) stays near zero in every bin.
        Panels (a)--(c) correspond to $A_{\rm CB}^{\rm input}=10^{-4}$, $10^{-5}$, and $4.4\times10^{-6}$.
        Points show the mean residual of the lensing bandpowers over 50 simulations relative to the realized input power, with standard errors, for joint (filled) and lensing-only (open) MUSE at $\Delta_P=1.4$ (circles) and $2.3\,\mu$K-arcmin (diamonds).}
    \label{fig:lensing_contamination_muse}
\end{figure}

We now look at the bias due to birefringence on the recovered lensing bandpowers.
We show the percentage lensing residuals in Fig.~\ref{fig:lensing_contamination_muse} for the lensing-only MUSE (open symbols) and joint MUSE (filled symbols).
The lensing-only case shows a clear positive bias that increases to small angular scales (high $L$), while the joint MUSE analysis does not.

We observe that neglecting birefringence biases the MUSE lensing bandpowers (open symbols in Fig.~\ref{fig:lensing_contamination_muse}).
The lensing-only case has differences that exceed $2\,\sigma$ for all six cases, with the biases largest at high $L$ and with more anisotropic birefringence power.
The magnitude of the bias is comparable at the two noise levels, although the error bars naturally shrink at lower noise.
The exception is $A_{\rm CB}^{\rm input}=10^{-4}$, where the bias is larger at the lower noise level.
At $\Delta_P=1.4\,\mu$K-arcmin, the peak biases are $(92.2\pm1.9)\%$, $(10.1\pm1.1)\%$, and $(5.1\pm1.1)\%$ for $A_{\rm CB}^{\rm input}=10^{-4}$, $10^{-5}$, and $4.4\times10^{-6}$, respectively, with uncertainties given by standard errors of the mean over the 50 realizations.
The positive bias arises because the lensing-only model attributes the Gaussian $B$-mode power added by the rotation to lensing.
The bias we find has the opposite sign from the negative bias found with a QE by Cai et al.~\citep{Cai2023Impact}, which also scales linearly with $A_{\rm CB}$.
The difference between the MUSE and QE biases is due to the $N^{(0)}$ bias subtraction step in the QE; the residual bias in the QE is dominated by connected terms that are negative at high $L$.

While the level of bias is small compared to statistical uncertainties on the small simulation fields, these levels of bias will become important for large-area CMB surveys.
At the BICEP/Keck upper limit~\citep{BICEPKeck2023XVII}, the lensing-only MUSE bias is about two-thirds of the single-realization scatter of the $1375<L<2000$ bin at $\Delta_P=1.4\,\mu$K-arcmin for our $164\,\mathrm{deg}^2$ patch.
The bias is not expected to decrease with more sky area, while the statistical uncertainties decrease as the inverse square root of the sky area.
Using this area scaling, the lensing bias would exceed $8\,\sigma$ for a CMB-S4-like wide survey covering 70\% of the sky~\citep{CMBS4Reference2019}.
Neglecting this potential bias might contaminate the cosmological inferences from the lensing measurement.

As shown by the filled symbols in Fig.~\ref{fig:lensing_contamination_muse}, this birefringence-induced bias is successfully mitigated with the joint MUSE framework.
The joint MUSE residuals are consistent with zero at all $L$ even with much more birefringence than is allowed by current measurements (i.e. at $A_{\rm CB} = 10^{-4}$).
The PTEs of the $\chi^2$ of these eight bins across the three input birefringence power levels range from 0.10 to 0.30.
The most significant residual in the $211<L<308$ bin appears to be due to sample variance in the set of simulations; similar deviations are seen in this bin in the lensing-only analysis that uses the same CMB and lensing maps with no birefringence.
We conclude that the joint MUSE framework has effectively mitigated the birefringence-induced bias on the recovered lensing power spectrum.

\subsection{Effect on lensing precision}
\label{lensing_precision}

\begin{figure}[!htbp]
    \centering
    \includegraphics[width=0.90\linewidth]{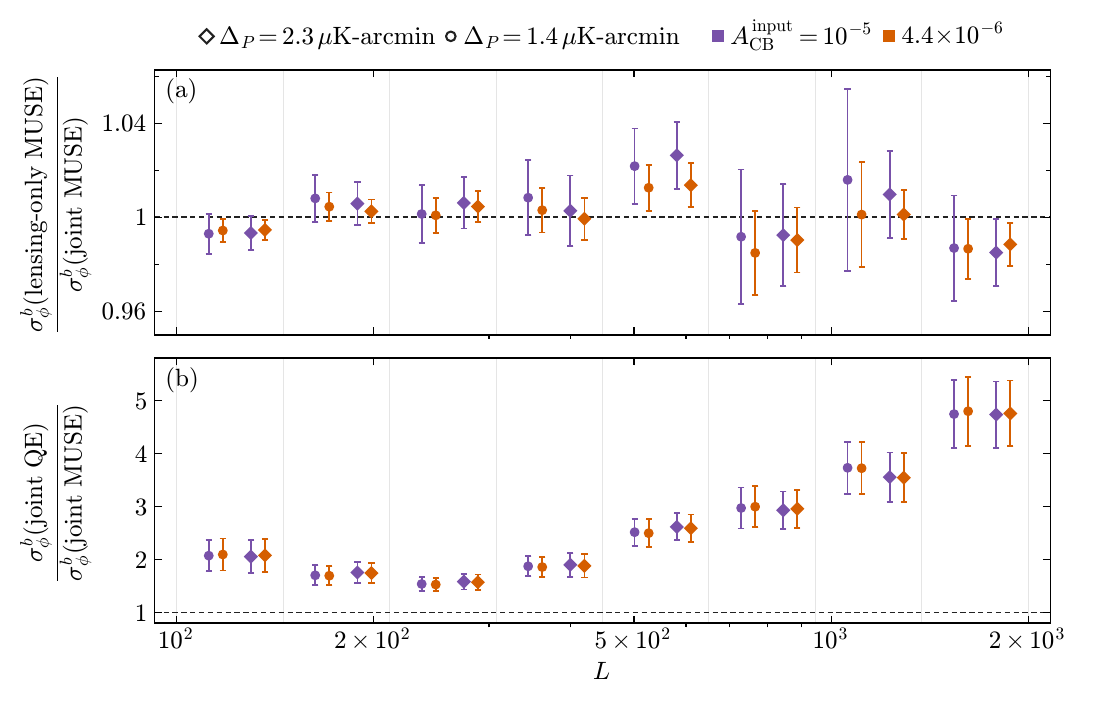}
    \caption{Simultaneously modeling anisotropic birefringence does not degrade the lensing precision at currently allowed levels of birefringence (a), and joint MUSE outperforms the joint QE at all angular scales, by a factor that grows towards high $L$ (b).
        Top panel (a): The ratio of lensing uncertainties for lensing-only to joint MUSE is consistent with unity.
        Error bars are bootstrap standard errors from 10,000 paired resamples of the 50 realizations.
        Bottom panel (b): The ratio of lensing uncertainties for the joint QE to joint MUSE is consistently above unity, and approaches five at high $L$.
        In both cases, the relative performance is largely unaffected by the assumed noise level or birefringence power.
        Note this breaks down in (a) for very large birefringence power ($A_{\rm CB}=10^{-4}$), which is not shown.
        The dashed black lines mark unity.
        Colors denote different levels of birefringence power $A_{\rm CB}^{\rm input}$, while circles versus diamonds denote the two noise levels $\Delta_P=1.4$ and $2.3\,\mu$K-arcmin.}
    \label{fig:lensing_precision}
\end{figure}

We find that simultaneously modeling birefringence does not appreciably degrade the lensing precision, and the joint MUSE estimator measures the lensing bandpowers much more precisely than the joint QE (Fig.~\ref{fig:lensing_precision}).
There is essentially no effect at $A_{\rm CB}\le 10^{-5}$; across the multipole bins, the ratios of single-realization scatter of lensing-only MUSE to joint MUSE range from $0.985\pm0.018$ to $1.026\pm0.014$ and are all consistent with unity.
This is possible because the two fields couple $E$ and $B$ modes with different angular dependencies.
As a result, the two responses are nearly orthogonal and the rotation is minimally degenerate with lensing.
While not plotted in Fig.~\ref{fig:lensing_precision}, we do see some effect on the lensing uncertainties at the extreme value of $A_{\rm CB}^{\rm input}=10^{-4}$, although we note this level of birefringence is ruled out by current observations.
For plausible levels of birefringence, we can simultaneously fit for both lensing and birefringence in MUSE without degrading the lensing measurement.

Across all bins at the two weaker amplitudes, the scatter of the joint QE is $1.53\pm0.13$ to $4.80\pm0.65$ times that of joint MUSE (Fig.~\ref{fig:lensing_precision}b).
The precision gain is smallest at $211<L<308$ and grows steadily towards high $L$ to between $3.5$ and $4.8$ at $L>946$.
This gain is nearly independent of $A_{\rm CB}$ and matches the no-birefringence case.

\subsection{Performance for birefringence}
\label{birefringence_amplitude}

\begin{figure}[!htbp]
    \centering
    \includegraphics[width=0.90\linewidth]{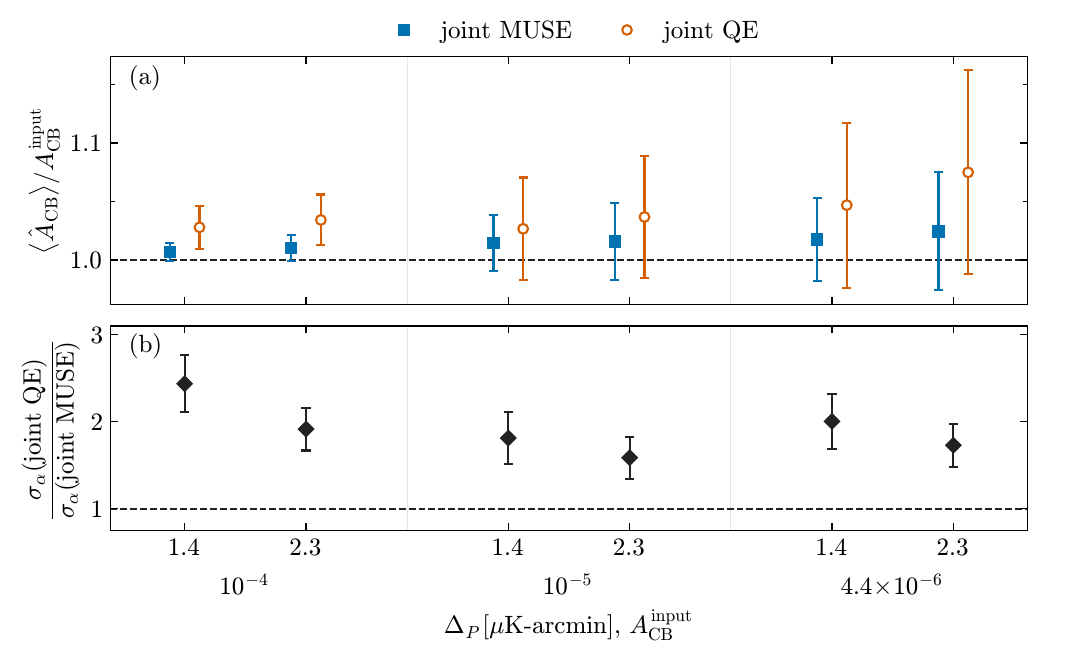}
    \caption{Both joint estimators recover $A_{\rm CB}$ without significant bias (a), and joint MUSE has a smaller uncertainty than the joint QE in all six configurations (b).
        Top panel (a): Mean recovered amplitude relative to the input for joint MUSE (squares) and joint QE (circles), with standard errors, for each of the six configurations.
        Bottom panel (b): Ratio of the single-realization scatter $\sigma_\alpha$ of $\hat A_{\rm CB}$ for joint QE to that for joint MUSE, with bootstrap standard errors.
        Dashed lines mark unity.}
    \label{fig:acb_recovery_precision}
\end{figure}

As shown in the top panel of Fig.~\ref{fig:acb_recovery_precision}, our simulations show that the joint MUSE and joint QE estimators recover $A_{\rm CB}$ without detectable bias.
The largest residual across the six configurations is $0.9\,\sigma$ for MUSE, entirely consistent with no bias.
The largest residual for the joint QE is larger at $1.6\,\sigma$, but still not clear evidence for a bias.

We show the relative precision of the birefringence measurement between the QE and MUSE in the bottom panel of Fig.~\ref{fig:acb_recovery_precision}.
We find that compared to the QE, MUSE reduces the $A_{\rm CB}$ measurement uncertainty by factors of $1.6\pm0.2$ to $2.4\pm0.3$.
This gain is slightly larger at the lower noise level at each input birefringence power level.
Joint MUSE outperforms joint QE across all six configurations.

\FloatBarrier

\section{Conclusions}
\label{conclusions}

We have demonstrated that while unmodeled anisotropic birefringence can bias CMB lensing power spectrum measurements, a joint-field analysis with MUSE can eliminate this bias without a significant penalty to the fidelity of the lensing reconstruction.
To do this, we extend the MUSE framework to jointly infer the CMB lensing bandpowers over $100\leq L\leq2000$ and the amplitude of a scale-invariant anisotropic birefringence spectrum.
We also consider the performance of the joint QE.
We test the ability of these three estimators to reconstruct the lensing power spectrum and birefringence power spectrum for four input birefringence power levels, from $A_{\rm CB}=0$ to $10^{-4}$, and two polarization noise levels of 1.4 and 2.3\,\ukarcmin.

We show that neglecting birefringence leads to a positive bias on the lensing power spectrum in the lensing-only MUSE results, with the fractional bias increasing with $L$.
The level of bias scales approximately linearly with the input birefringence power, $A_{\rm CB}$.
For $A_{\rm CB}^{\rm input}=4.4\times10^{-6}$ (the BICEP/Keck 95\% upper limit~\citep{BICEPKeck2023XVII}) and 1.4\,\ukarcmin{} noise, the bias in a lensing-only MUSE analysis reaches $(5.1\pm1.1)\%$ at $1375<L<2000$.
This would lead to a $\gtrsim8\,\sigma$ bias in the lensing reconstruction at these angular scales for a CMB-S4-like survey of 70\% of the sky.
The joint MUSE analysis removes this bias, leaving residuals consistent with zero even at $A_{\rm CB}^{\rm input}=10^{-4}$ which is $37\times$ existing limits on cosmic birefringence from~\citep{coerver26}.

Importantly, the joint MUSE analysis removes the birefringence-induced bias without significantly degrading the lensing precision.
Even for cosmic birefringence power that is a factor of a few larger than current limits, the precision of the lensing reconstruction with the joint MUSE framework is within 3\% of the lensing-only case.
We find the joint MUSE estimator outperforms the joint QE, as one would expect.

Finally, both joint MUSE and joint QE recover the birefringence amplitude without detectable bias across all six configurations.
As with lensing, the precision of the birefringence reconstruction is improved with MUSE.
For $A_{\rm CB}^{\rm input}=4.4\times10^{-6}$ and the assumed noise levels, the MUSE estimator reduces the empirical uncertainty on $A_{\rm CB}$ by a factor of $\sim$\,1.8 relative to the joint QE.

Given the promising performance demonstrated in our simulations, a natural next step is to extend the joint framework to more realistic survey conditions before applying it to data.
In particular, survey masks, which introduce incomplete sky coverage and the resulting mode coupling, and residual polarized foregrounds, which may bias the inferred lensing and birefringence signals, will need to be addressed.
These developments would prepare the framework for application to polarization data from SPT-3G~\citep{Ge2025Cosmology}, SPT-3G+~\citep{spt-3g+} and the Simons Observatory~\citep{SimonsObservatory2019} as ultra-low-noise CMB maps across wide areas of sky become available.

\acknowledgments

RZ acknowledges support from ChatGPT and Claude in code development and language editing.
All output was reviewed and verified by the authors.
We acknowledge support from the Australian Research Council's Discovery Projects scheme (DP260100705).
This research was supported by The University of Melbourne's Research Computing Services and the Petascale Campus Initiative.

\bibliographystyle{JHEP}
\bibliography{main}

\end{document}